\documentclass[a4paper,11pt]{article}
\usepackage[dvips]{graphicx}
\usepackage{cite}

\begin{document}

\title{Fluorescence energy transfer enhancement in aluminum nanoapertures}

\date{}

\author{Juan de Torres \thanks{CNRS, Aix-Marseille Universit\'e, Centrale Marseille, Institut Fresnel, UMR 7249, 13013 Marseille, France}, Petru Ghenuche \footnotemark[1], Satish Babu Moparthi \footnotemark[1],\\ Victor Grigoriev \footnotemark[1] \thanks{Current address: CSIRO Energy Centre, Mayfield West, NSW 2304 Australia}, J\'{e}r\^{o}me Wenger \footnotemark[1] \thanks{Corresponding author: jerome.wenger@fresnel.fr}}

\maketitle

\section*{Abstract}

Zero-mode waveguides (ZMWs) are confining light into attoliter volumes, enabling single molecule fluorescence experiments at physiological micromolar concentrations. Among the
fluorescence spectroscopy techniques that can be enhanced by ZMWs, F\"{o}rster resonance energy transfer (FRET) is one of the most widely used in life sciences. Combining zero-mode waveguides with FRET provides new opportunities to investigate biochemical structures or follow interaction dynamics at micromolar concentration with single molecule resolution. However, prior to any quantitative FRET analysis on biological samples, it is crucial to establish first the influence of the ZMW on the FRET process. Here, we quantify the FRET rates and efficiencies between individual donor-acceptor fluorophore pairs diffusing in aluminum zero-mode waveguides. Aluminum ZMWs are important structures thanks to their commercial availability and the large literature describing their use for single molecule fluorescence spectroscopy. We also compare the results between ZMWs milled in gold and aluminum, and find that while gold has a stronger influence on the decay rates, the lower losses of aluminum in the green spectral region provide larger fluorescence brightness enhancement factors. For both aluminum and gold ZMWs, we observe that the FRET rate scales linearly with the isolated donor decay rate and the local density of optical states (LDOS). Detailed information about FRET in ZMWs unlocks their application as new devices for enhanced single molecule FRET at physiological concentrations.

\section*{Introduction}

F\"{o}rster resonance energy transfer (FRET) is one of the most widely used techniques in single molecule studies applied to life sciences \cite{Ha08}. FRET involves the nonradiative transfer of electronic excitation energy from an excited donor D$^*$ to a ground-state acceptor molecule A. The energy transfer efficiency goes down with the inverse sixth power of the D-A distance, which makes FRET highly sensitive to the relative position of donor and acceptor fluorophores at the nanoscale. Therefore, FRET is often used to quantify the spatial relationship between two fluorophore-labeled sites in biological structures \cite{Deniz99,Lilley08}, study conformational changes in macromolecules \cite{Weiss00,Schuler02} or follow interaction dynamics between proteins, DNA, RNA and peptide molecules \cite{Schuler08,Medintz03}.

FRET experiments are generally implemented on a diffraction-limited confocal microscope which requires pico to nanomolar concentrations so as to isolate a single molecule in the detection volume. To go beyond the restrictions imposed by diffraction, metal nanoapertures (also known as zero-mode waveguides ZMWs) are powerful tools for monitoring real-time dynamics of single molecules at micromolar concentrations \cite{Levene03,Craighead08}. By confining the illumination light to the bottom of the subwavelength aperture, detection volumes down to the attoliter range are commonly reached, reducing the detection volume by three orders of magnitude as compared to diffraction-limited confocal microscopy \cite{TinnefeldRev13,PunjWires14}. Within ZMWs, experiments can be performed at the physiological concentration of molecules as found in living cells. ZMWs were successfully used on various investigations such as single-molecule DNA and tRNA sequencing \cite{Eid09,Uemura10,Meller10}, protein-protein interaction dynamics \cite{Samiee05,Miyake08}, and cell membrane receptors organization \cite{WengerMemb07,Moran07,Richards12,Kelly14}.

The use of ZMWs appears promising to extend single molecule FRET towards higher physiological concentrations. However, prior to any quantitative FRET analysis, it is crucial to establish first the influence of the ZMW on the FRET process. The question arises as FRET naturally competes with the donor radiative emission and the donor nonradiative energy losses to the metal structure, which are both known to be influenced by the photonic environment through the local density of optical states (LDOS) \cite{Drexhage70,Novotnybook}. The LDOS is commonly defined as the number of electromagnetic modes per unit volume and frequency at the position of the dipole emitter where the energy can be released during the spontaneous emission process \cite{Novotnybook}. The energy release can occur by photon radiation or by local absorption and non-radiative transitions. The LDOS is therefore proportional to the emitter's total decay rate (inverse of the fluorescence lifetime), including both radiative and non-radiative transitions.

Several earlier works have considered the influence of photonic nanostructures on the FRET process: some conclude that the FRET rate depends linearly on the donor emission rate and the LDOS \cite{Andrew00,Carminati,Enderlein,Bradley11,Bradley14}, while some others report a FRET rate independent of the LDOS \cite{Polman05,Blum12,Rabouw14,Meixner14}. Earlier experiments on aluminum C-shaped nanoapertures did not reveal noticeable changes of the FRET efficiency \cite{Fore07}. A similar trend was reported for circular aluminum ZMWs but without quantitative data \cite{Zhao14}, while another study quantified a reduction of the FRET efficiency by 15\% induced by the ZMWs \cite{Puglisi14}. Recently, our group has investigated FRET in ZMWs milled in gold films, reported a linear dependence of the FRET rate on the LDOS and quantified a slight variation of the FRET efficiency with the aperture diameter \cite{Ghenuche14}.

Here, we build on our previous methodology \cite{Ghenuche14} to investigate the role of aluminum ZMWs on the FRET process between individual donor-acceptor fluorophore pairs on double stranded DNA linkers. ZMWs milled in aluminum are by far the most widely used structures as compared to other metals such as gold or copper \cite{Levene03,Craighead08,TinnefeldRev13,PunjWires14,Eid09,Uemura10,Meller10,Samiee05,Miyake08,WengerMemb07,Moran07,Richards12,Kelly14,Fore07,Zhao14,Puglisi14}, thanks to their commercial availability, low chemical reactivity and the availability of surface passivation protocols \cite{Korlach08}. Due to the large interest raised by aluminum ZMWs and FRET separately, we believe that a clear quantification of FRET in aluminum ZMWs is necessary. Moreover, to provide a complete picture and discuss the role of plasmonic effects, we compare the results found with gold and aluminum, and find some significant differences on the fluorescence brightness and the fluorescence lifetime reduction notably.

A key strength of our approach is the use of two different independent characterization methods to measure the FRET rate $\Gamma_{FRET}$ and efficiency $E_{FRET}$. This is made possible by monitoring simultaneously both the donor and the acceptor fluorescence photodynamics. The first method is based on the donor lifetime reduction in the presence of the acceptor. The second method is based on fluorescence burst intensity analysis accounting for the acceptor fluorescence increase in presence of the donor and the donor fluorescence quenching in presence of the acceptor. Both methods converge remarkably well towards similar results, as we show in Fig.~6 of the Supporting Information. The FRET efficiency is defined as the probability of energy transfer over all donor transition events: $E_{FRET}=\Gamma_{FRET}/\Gamma_{DA}=\Gamma_{FRET}/(\Gamma_{FRET}+\Gamma_{D})$, where $\Gamma_{D}=\Gamma_{D,rad}+\Gamma_{D,nonr}$ is the total decay rate of the isolated donor (accounting for both radiative and non-radiative transitions) and $\Gamma_{DA}=\Gamma_{D}+\Gamma_{FRET}$ is the total decay rate of the donor in the presence of the acceptor. As demonstrated for instance in \cite{Novotnybook}, the LDOS is proportional to the isolated donor total decay rate $\Gamma_D$.

\begin{figure}[h!]
\begin{center}
\includegraphics{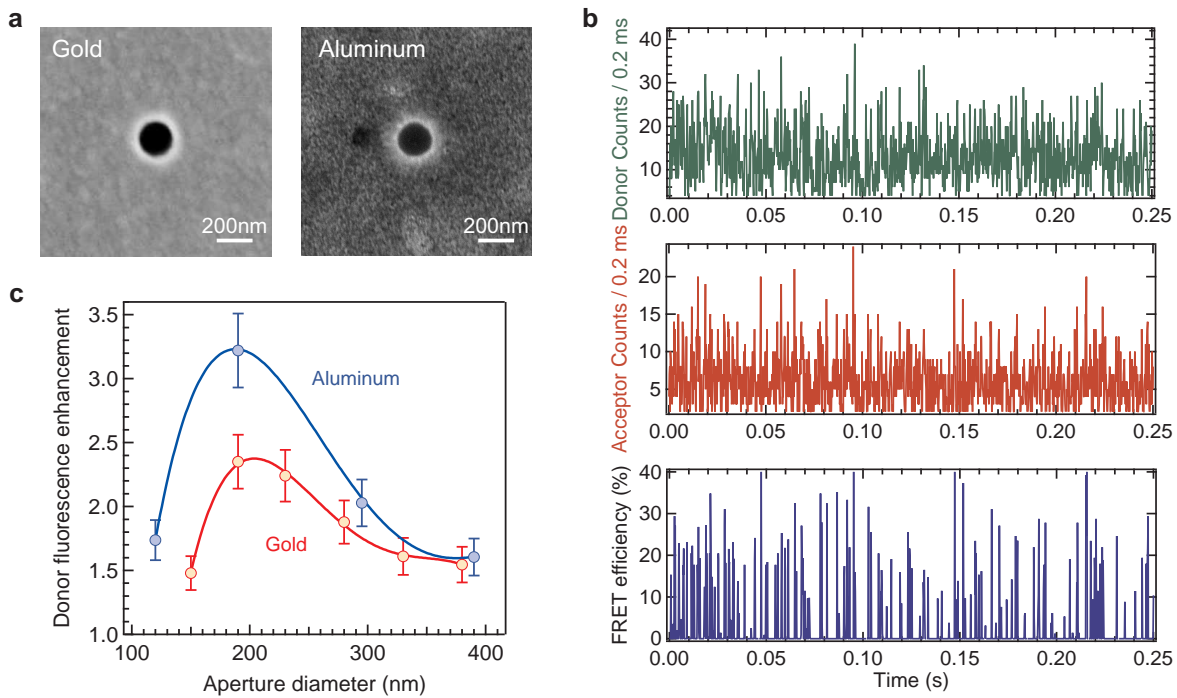}
\caption{(a) Scanning electron microscopy image of nanoapertures of 190~nm diameter milled in gold or aluminum with focused ion beam. (b) Typical fluorescence time traces for Atto550 donor and Atto647N acceptor diffusing in an aluminum nanoaperture of 190~nm diameter. The donor and acceptor molecules are fixed on a double stranded DNA linker with a donor-acceptor separation of 10.2~nm (30 base pairs). For each fluorescence burst exceeding the detection threshold in the donor or acceptor detection channel, a FRET efficiency is calculated (bottom trace). The binning time is 0.2~ms which is similar to the diffusion time of DNA samples in the ZMW (see Supporting Information Fig.~7). The total trace duration is 200~s. (c) Fluorescence brightness enhancement factors for the isolated donor for gold and aluminum apertures. Lines are guide to the eyes.} \label{FigIntro}
\end{center}
\end{figure}

\section*{Results and Discussion}

In this work, we use circular nanoapertures with diameters from 110 to 400~nm milled by focused ion beam in a 150~nm thick aluminum or gold layer deposited on a glass microscope coverslip (Fig.~\ref{FigIntro}a). The FRET samples are provided by Atto550 donor and Atto647N acceptor covalently attached on double stranded DNA molecules with different distances between the donor and acceptor. To ease the discussion, we focus on two D-A separations of 20 base pairs ($\sim 6.8$~nm) and 30 base pairs ($\sim 10.2$~nm). The ZMWs are covered by a solution containing the FRET donor-acceptor pairs at around one micromolar concentration. The experiments monitor the fluorescence bursts generated as the fluorescent molecules cross the ZMW volume. Both the donor and the acceptor emission photodynamics are recorded simultaneously with picosecond resolution in time-tagged time-resolved (TTTR) mode, resulting in fluorescence time traces as displayed in Fig.~\ref{FigIntro}b. We have also checked that our conclusions are concentration invariant, see Fig.~8 of the Supporting Information.

First we consider the effect of the ZMW on the fluorescence brightness of the isolated donor. Section~2 of the Supporting Information details the use of fluorescence correlation spectroscopy (FCS) to quantify the fluorescence brightness enhancement. Typical FCS correlation traces are shown in Fig.~7. We measure a typical fluorescence brightness enhancement of 3.2 in a 190~nm aluminum ZMW, while the fluorescence enhancement is only 2.35 in a gold ZMW of similar dimensions (Fig.~\ref{FigIntro}c). These results stand in good agreement with numerical simulations \cite{Blair10,Blair07} and the behavior expected from previous measurements on red fluorescent dyes \cite{Davy08,Wenger08}. The difference between the fluorescence enhancement factors for aluminum and gold ZMWs stems mainly from the 550~nm laser excitation intensity which is lower in gold ZMW due to the increased losses for gold permittivity at 550~nm \cite{Blair10,Blair07,Rakic,RefractiveIndex} (the phenomena leading to the observation of enhanced fluorescence are discussed with more details below). From the fluorescence intensity enhancement, it seems that aluminum ZMWs are therefore preferable over gold for excitation wavelengths below 570~nm. Hereafter, we will investigate the influence of the metal on the fluorescence lifetimes and FRET rates in ZMWs.

\begin{figure}[t!]
\begin{center}
\includegraphics{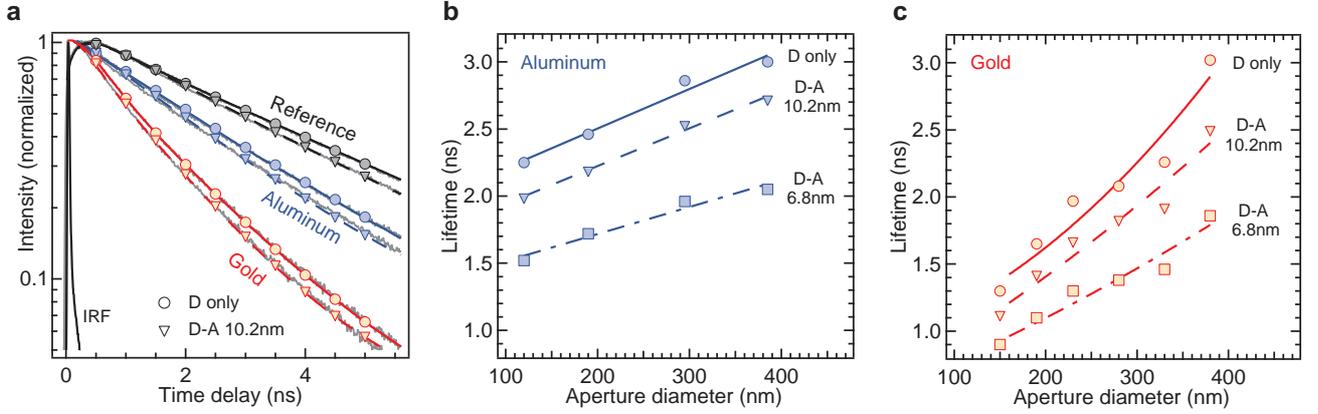}
\caption{(a) Normalized donor fluorescence decay traces when no acceptor is present (circles) or when an acceptor is set at 10.2~nm distance (triangles) for the confocal reference and for a 190~nm aperture milled in aluminum or gold. Lines with symbols are numerical fits convoluted by the instrument response function (IRF). (b) Donor fluorescence lifetime as function of the aperture diameter for ZMWs milled in aluminum and for different donor-acceptor separations. Lines are guide to the eyes. (c) Same as (b) for apertures milled in gold. The reference lifetimes for the confocal setup are: 3.67~ns for the isolated donor, 3.13~ns for the donor with acceptor set at 10.2~nm, and 2.44~ns for the donor with acceptor set at 6.8~nm. All lifetimes are listed in the Supporting Information table S2.} \label{FigLifetime}
\end{center}
\end{figure}

For each experiment, we construct the donor fluorescence decay trace by time-correlated single photon counting. Figure~\ref{FigLifetime}a shows typical decay traces for the donor in confocal setup and in gold or aluminum ZMWs of 190~nm diameter. Two effects are readily seen on the raw decay traces. First, the donor emission dynamics are faster in the gold ZMW than in aluminum or confocal, as a consequence of the stronger influence of gold on the LDOS. Second, in all cases the donor emission dynamics are further accelerated by the acceptor presence, which opens a new decay channel by FRET. Consequently, the fluorescence lifetime of the donor in presence of the acceptor $\tau_{DA}$ is shorter than the lifetime of the isolated donor $\tau_{D}$. Numerical interpolation of the decay traces (see Experimental Section) quantifies the average donor fluorescence lifetimes for aluminum (Fig.~\ref{FigLifetime}b) and gold ZMWs (Fig.~\ref{FigLifetime}c). A clear reduction of the donor emission lifetime (donor quenching) is observed as the D-A distance is reduced, with a similar trend observed for all nanoaperture diameters (Table~S2 of the Supporting Information provides a list of all fluorescence lifetimes measured). We point out that an intrinsic limitation of our approach based on random diffusion is that it provides only the spatially averaged fluorescence lifetime, while lifetime variations are expected within the ZMW depending on the position and orientation of the emitter respective to the metal \cite{Heucke14,Pibiri14}. Nevertheless, our net results remain relevant to describe the average properties of molecules randomly positioned inside ZMWs or randomly diffusing \cite{PunjWires14,Fore07,Zhao14,Puglisi14}.

\begin{figure}[t!]
\begin{center}
\includegraphics{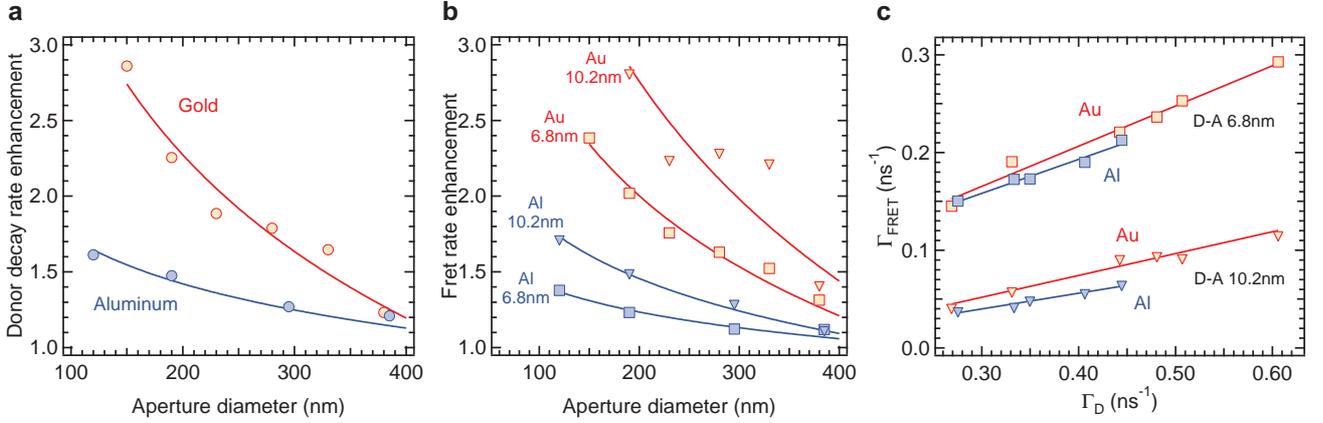}
\caption{(a) Comparison of the decay rate enhancement (lifetime reduction) for the isolated donor as function of the ZMW diameter for gold and aluminum layers. (b) FRET rate enhancement as function of the ZMW diameter for different metals (aluminum: blue; gold: red) and donor-acceptor distance (6.8~nm: squares; 10.2~nm: triangles). (c) FRET rate $\Gamma_{FRET}$ as function of the isolated donor decay rate $\Gamma_D$ for two donor-acceptor separations in ZMWs milled in aluminum (blue markers) or gold (red markers).} \label{FigEnhFRET}
\end{center}
\end{figure}

The fluorescence lifetime $\tau_D$ of the isolated donor decreases as the ZMW diameter is reduced, bringing the metallic walls closer to the dye (Fig.~\ref{FigLifetime}b,c). Alternatively, the lifetime reduction can be seen as an increase of the isolated donor total decay rate $\Gamma_D=1/\tau_D=\Gamma_{D,rad}+\Gamma_{D,nonr}$, which accounts for both radiative and non-radiative transitions. Following the common definition in nanophotonics \cite{Novotnybook}, $\Gamma_D$ is proportional to the LDOS at the donor emission wavelength. Figure~\ref{FigEnhFRET}a compares the enhancement of the isolated donor decay rate $\Gamma_D$ for aluminum and gold ZMWs. A clear difference is seen, with gold ZMWs providing $\sim 2\times$ faster photodynamics and higher LDOS, in agreement with numerical computations \cite{Davy08} and the presence of stronger plasmonic resonances for gold.

In the presence of the acceptor, the total decay rate of the donor is increased by the additional FRET decay channel $\Gamma_{DA}=\Gamma_{D}+\Gamma_{FRET}$. The lifetime data in Fig.~\ref{FigLifetime}b,c thus quantifies the FRET rate as $\Gamma_{FRET}=\Gamma_{DA}-\Gamma_{D}=1/\tau_{DA}-1/\tau_{D}$. For both aluminum and gold ZMWs, we observe similar increases of the FRET rate $\Gamma_{FRET}$ and the isolated donor rate $\Gamma_{D}$ (Fig.~\ref{FigEnhFRET}a,b), with a larger rate enhancement in the case of gold. Remarkably, our data indicates that the FRET rate $\Gamma_{FRET}$ scales linearly with the isolated donor rate $\Gamma_{D}$ for both metals and D-A distances (Fig.~\ref{FigEnhFRET}c). In the case of aluminum, the influence of the ZMW on the LDOS is less pronounced, and consequently the range of donor rates $\Gamma_{D}$ being probed is smaller than for gold. Nonetheless, the FRET rate in aluminum ZMWs follows a clear linear dependence with $\Gamma_{D}$, confirming the possibility to control FRET with nanophotonics \cite{Andrew00,Ghenuche14}.

Another remarkable feature of FRET in ZMWs is the larger FRET rate enhancement in the case of increased D-A separations (Fig.~\ref{FigEnhFRET}b). For short D-A distances (on the order of the F\"{o}rster radius or below), the direct dipole-dipole energy transfer dominates and the LDOS has a moderate effect on the FRET rate, especially for nanophotonic structures of the size of a half wavelength with limited field gradients \cite{Meixner14}. However, for larger D-A distances and structures with more pronounced field confinement, a supplementary contribution from the energy transfer mediated by the nanostructure can further enhance the apparent FRET rate \cite{Carminati,Ghenuche14}.

\begin{figure}[t!]
\begin{center}
\includegraphics{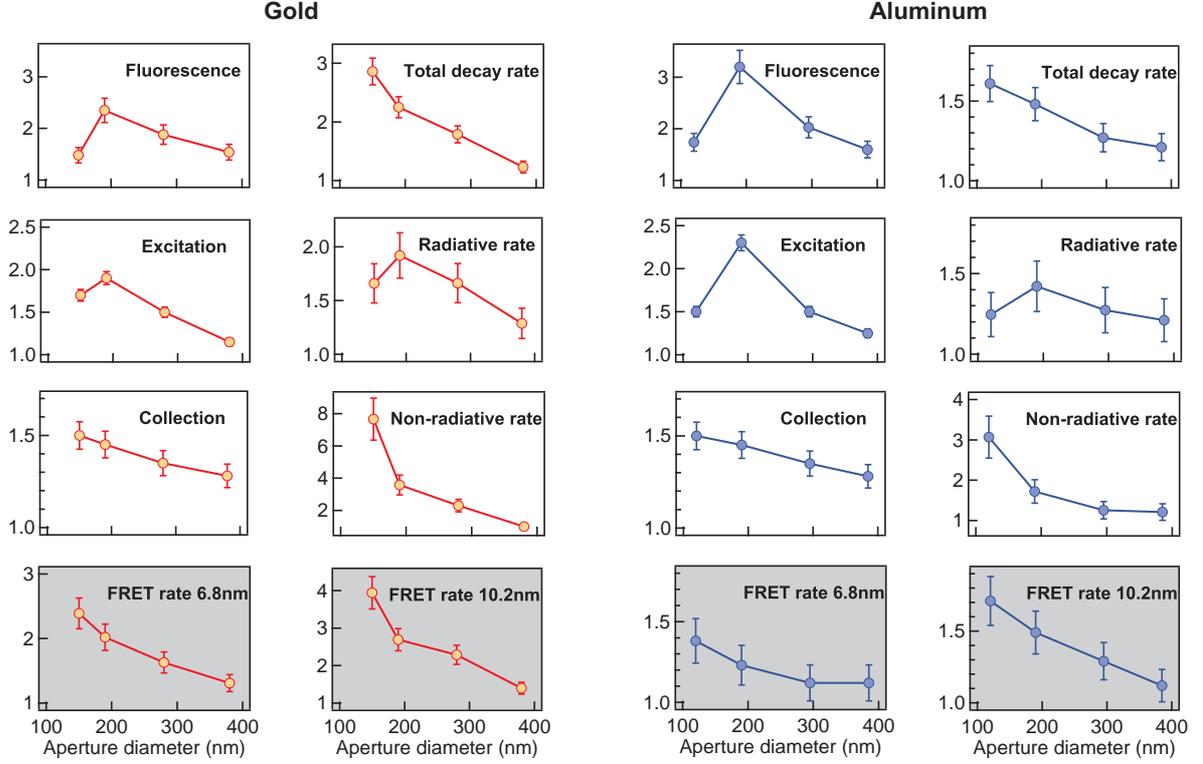}
\caption{Summary of the enhancement factors for gold and aluminum ZMWs. The total decay rate is for the isolated donor (no FRET). For both metals and D-A distances, the FRET rate enhancement evolves as the total decay rate of the isolated donor. The gains always refer to the confocal reference. See Experimental Section for details on the computation procedure.} \label{FigComp}
\end{center}
\end{figure}

To summarize the differences between aluminum and gold ZMWs, we quantify the different effects leading to enhanced photodynamics (Fig.~\ref{FigComp} and Fig.~9 of the Supporting Information). As already seen in Fig.~\ref{FigIntro}c, the fluorescence brightness appears significantly larger for aluminum than for gold. As a consequence of the increased losses for gold at the 550~nm illumination wavelength \cite{Rakic,RefractiveIndex}, the excitation intensity enhancement is smaller for gold than for aluminum, which contributes to the smaller apparent fluorescence enhancement found for gold. Thanks to the stronger plasmon response of gold and probably also because of the oxide layer naturally present at the surface of aluminum, the radiative rate and total decay rate enhancements are $\sim 2\times$ larger for gold. For comparison, we add on Fig.~\ref{FigComp} the values found for the FRET rate enhancement in the cases of 6.8 and 10.2~nm D-A distances. For both metals, the FRET rate enhancement is found in agreement with the isolated donor total decay rate. We also find that the FRET rate enhancement does not scale with the excitation intensity gain (as one would expect for FRET) nor with the total fluorescence brightness enhancement of the isolated donor. The case of the 190~nm aluminum ZMW is particularly illustrative: the fluorescence enhancement and local excitation intensity are high, but not the donor decay rate or the FRET rate.

\begin{figure}[t!]
\begin{center}
\includegraphics{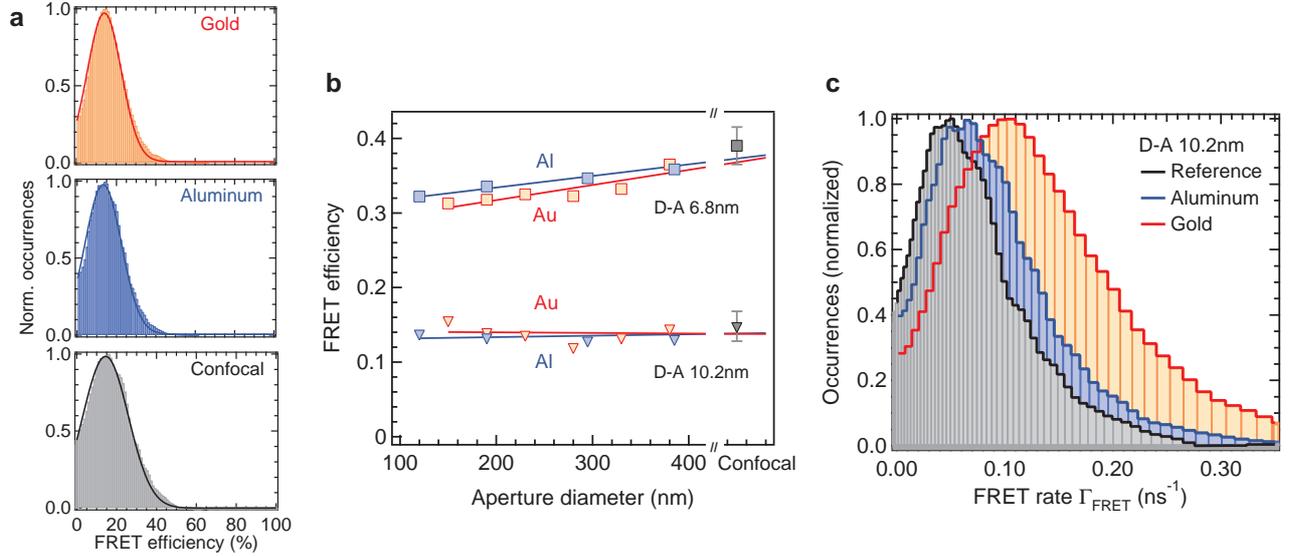}
\caption{(a) FRET efficiency histograms extracted from fluorescence burst analysis for 10.2~nm donor-acceptor separation in 190~nm nanoaperture milled in gold or aluminum. The FRET histogram found for the confocal reference is also displayed for comparison. Lines are Gaussian fits used to determine the center FRET efficiency. (b) FRET efficiency $E_{FRET}$ as function of the aperture diameter for different metals (aluminum: blue markers; gold: red markers) and donor-acceptor distance (6.8~nm: squares; 10.2~nm: triangles). The grey markers represent the values for the confocal reference. Lines are linear fits. (c) Distribution of FRET rates deduced from the efficiency histograms in (a) and the average isolated donor lifetime in Fig.~\ref{FigLifetime}b,c for 190~nm apertures and 10.2~nm donor-acceptor separation.} \label{FigEffic}
\end{center}
\end{figure}

Complementary to the fluorescence lifetime analysis, our procedure also records the fluorescence bursts intensities from which the FRET efficiency $E_{FRET}$ can be computed and collected in a histogram. The FRET efficiency is defined as the probability of energy transfer over all donor transition events: $E_{FRET}=\Gamma_{FRET}/\Gamma_{DA}=\Gamma_{FRET}/(\Gamma_{FRET}+\Gamma_{D})$, where $\Gamma_{D}=\Gamma_{D,rad}+\Gamma_{D,nonr}$ is the total decay rate of the isolated donor (without acceptor, accounting for both radiative and non-radiative transitions). The Experimental Section details the procedure to take into account the direct excitation of the acceptor by the laser light, donor emission crosstalk into the acceptor channel, and differences in the quantum yields and detection efficiencies of the donor and acceptor emission. Figure~\ref{FigEffic}a summarizes the FRET efficiency histograms for 10.2~nm FRET pairs in gold and aluminum ZMWs of 190~nm diameter and in the confocal reference case (non-normalized histograms are provided in the Supporting Information Fig.~9). Although the LDOS is clearly different between the three cases (Fig.~\ref{FigLifetime}a), all histograms have similar average values and widths. The trend is further evidenced by recording the average FRET efficiency as the ZMW diameter is varied (Fig.~\ref{FigEffic}b). For the 6.8~nm D-A separation, we observe a small reduction of 6\% of the FRET efficiency when the ZMW diameter goes down to 110~nm. The results for both aluminum and gold ZMW follow the same trend. For the 10.2~nm D-A separation, the FRET efficiency is almost constant for all ZMW diameters. The small 2\% offset seen between gold and aluminum ZMWs in this case is an artefact related to the different sample solution in use, as confirmed by confocal measurements. Additionally, the FRET histograms can be used to compute the statistical distribution of the FRET rates $\Gamma_{FRET}= \Gamma_{D} \,\, E_{FRET}/(1-E_{FRET})$ using the separate measurements of the average isolated donor decay rates $\Gamma_{D}$  (Fig.~\ref{FigLifetime}a) and the FRET efficiency from burst analysis (Fig.~\ref{FigEffic}a). This leads to the distributions seen in Fig.~\ref{FigEffic}c, which illustrates the enhancement of the FRET rate $\Gamma_{FRET}$ brought by the ZMW.

\section*{Conclusions}

We have detailed the influence of aluminum ZMWs on the FRET process between individual donor-acceptor fluorophore pairs. We have measured separately the FRET rate and the FRET efficiency by measuring simultaneously the donor and the acceptor fluorescence photodynamics. We have found that both the FRET rate and FRET efficiency are consistent with the linear dependence of the FRET rate on the LDOS in ZMWs nanoapertures, and that accordingly, the FRET efficiency is marginally affected by the ZMW.
We have also compared the results obtained with gold and aluminum ZMWs, and found some significant differences on the fluorescence brightness and the fluorescence lifetime reduction notably. While aluminum ZMWs have a lower influence on the decay rates than gold ZMWs of similar diameter, the lower losses of aluminum at 550~nm wavelength enable higher gains in local excitation intensity and fluorescence brightness. To compare between metals and find the most suitable ZMW material for a chosen illumination wavelength, the ratio $-\mathrm{Re}(\varepsilon)/\mathrm{Im}(\varepsilon)$ of the real part over the imaginary part of the complex permittivity $\varepsilon$ is often used as a figure of merit \cite{West,Jackson}. While gold has a high figure of merit in the red spectral region, aluminum performs better for illumination wavelengths below 560~nm \cite{Rakic,RefractiveIndex} (Fig.~13 of the Supporting Information). We anticipate that this quantification of the FRET process in ZMWs will promote their application as new devices for enhanced single molecule FRET analysis at physiological micromolar concentrations \cite{Levene03,TinnefeldRev13,PunjWires14,Fore07,Zhao14,Puglisi14}.

\section*{Experimental Section}

\textbf{Nanoaperture ZMW fabrication.} Nanoapertures are milled by focused ion beam (FEI Strata Dual Beam 235) on 150~nm thick aluminum or gold films deposited using thermal evaporation on standard 150~$\mu$m thick microscope glass coverslips.

\textbf{DNA samples}. Double-stranded DNA constructs of 51 base pairs length are designed with one Atto550 donor on the forward strand, and one Atto647N acceptor on the reverse strand. Fluorescently labeled and HPLC-purified DNA single strands are obtained from IBA (G\"{o}ttingen, Germany), modified with the corresponding N-hydroxysuccinimidyl ester (NHS) donor and acceptor fluorophore derivatives of ATTO550 and ATTO647N. Fluorophores are covalently linked to an amino-C6-modified thymidine with NHS-chemistry via base labeling. The forward strand sequence is

5'~CCTGAGCGTACTGCAGGATAGCCTATCGCGTGTCATATGCTGT$\mathrm{\mathbf{T_D}}$CAGTGCG~3'.

The reverse strand sequence is

5'~CGCACTGAACAGCATATGACACGCGA$\mathrm{\mathbf{T_{20}}}$AGGCTATCC$\mathrm{\mathbf{T_{30}}}$GCAGTACGCTCAGG~3'.

The distances between fluorescent labels are set so that the donor and acceptor are separated by 20 or 30 base pairs (corresponding to $\sim6.8$ and 10.2~nm separations respectively). The characteristic F\"{o}rster radius computed for Atto550 and Atto647N in pure water is 6.5~nm. The reference sequences carrying only the isolated donor or acceptor are constructed with unlabeled complementary strand respectively. The strands are annealed at 10~$\mu$M concentration in 20~mM Tris, 1~mM EDTA, 500~mM NaCl, 12~mM MgCl$_2$ buffer, and by heating to 95$^{\circ}$C for 5 min followed by slow cooling to room temperature. Double stranded DNA stocks are diluted in a 10~mM Hepes-NaOH buffer, pH~7.5 (Sigma-Aldrich).

\textbf{Experimental setup.} Experiments are performed on a confocal inverted microscope with a Zeiss C-Apochromat 63x 1.2NA water-immersion objective. The excitation is provided by a iChrome-TVIS laser (Toptica GmbH) delivering 3~ps pulses at 40~MHz repetition rate and 550~nm wavelength. We use on average 40~$\mu$W excitation power to avoid saturating the fluorescent dyes. Filtering the laser excitation is performed by a set of two bandpass filters (Chroma ET525/70M and Semrock FF01-550/88). Dichroic mirrors (Chroma ZT594RDC and ZT633RDC) separate the donor and acceptor fluorescence from the reflected laser light. The detection is performed by two avalanche photodiodes (Micro Photon Devices MPD-5CTC with $<50$~ps timing jitter) with $620 \pm 20$~nm (Chroma ET605/70M and ET632/60M) and $670 \pm 20$~nm (Semrock FF01-676/37) fluorescence bandpass filters for the donor and acceptor channels respectively. The photodiode signal is recorded by a fast time-correlated single photon counting module (Hydraharp400, Picoquant GmbH) in time-tagged time-resolved (TTTR) mode. Each trace duration is typically of 200s. The temporal resolution for fluorescence lifetime measurements is 37~ps at half-maximum of the instrument response function.

\textbf{Fluorescence lifetime analysis.} The time correlated single photon counting (TCSPC) histograms are fitted using Levenberg-Marquard optimisation, implemented  using the commercial software Symphotime 64 (Picoquant GmbH). The model considers a single exponential decay reconvoluted by the instrument response function (IRF, as shown on Fig.~\ref{FigLifetime}a). The time interval for fit is set to ensure that more than 85\% of the detected count events are taken into account in the region of interest. As shown already in some of our earlier work \cite{Ghenuche14,Wenger08,Aouani11} and by the results in Fig.~\ref{FigLifetime}a, a single exponential fit is a satisfactory approximation to the fluorescence decay traces in nanoapertures (Fig.~11 of the Supporting Information). For gold aperture diameters below 250~nm, a background signal is detected due to the metal photoluminescence, which we take into account by adding a supplementary decay term with a fixed 5~ps characteristic time (see Fig.~11 of the Supporting Information).

\textbf{FRET efficiency analysis.} For every detected fluorescence burst above the background noise, we record the number of photons in the acceptor channel $n_a$ and in the donor channel $n_d$. For both the confocal and the ZMW configurations, the threshold for burst recognition is set to the sum of the mean plus one standard deviation of the summed trace of donor and acceptor channels. To avoid experimental artifacts in the FRET analysis, we also carefully characterize the optical response of the isolated donor and the isolated acceptor for each ZMW. In the case of the isolated donor, we measure the fraction $\alpha$ of photons from the donor that fall into the acceptor detection channel due to non-negligible spectral overlap between the donor emission and the acceptor detection window. For all our measurements, we find a constant $\alpha=0.17$ that is not affected by the ZMW. In the case of the isolated acceptor, we record the number $n_{ao}^{de}$ of photons that result from the direct excitation of the acceptor dye by the laser light. The FRET efficiency is then computed:
\begin{equation}
    E_{FRET} = \frac{n_a - \alpha n_d - n_{ao}^{de} }{n_a - \alpha n_d - n_{ao}^{de} + \gamma n_d}
\end{equation}
Here $\gamma=\eta_a \phi_a / \eta_d \phi_d$ accounts for the differences in quantum yields ($\phi_a$ and $\phi_d$) and fluorescence detection efficiencies ($\eta_a$ and $\eta_d$) between the acceptor and donor. For the confocal reference and the ZMWs, we estimate $\gamma=1.3$ in the case of our setup. The full trace analysis is implemented using the software Symphotime 64 (Picoquant GmbH).

\textbf{Quantification of the enhancement factors.} The fluorescence enhancement $\eta_F$ for the isolated donor is quantified by burst intensity analysis or alternatively FCS (Fig.~\ref{FigIntro}c) \cite{Davy08,Wenger08}. Below the fluorescence saturation regime, three phenomena contribute to the fluorescence enhancement $\eta_F$: the gains in collection efficiency $\eta_{\kappa}$, quantum yield $\eta_{\phi}$, and excitation intensity $\eta_{exc}$, so that $\eta_F = \eta_{\kappa} \eta_{\phi} \eta_{exc}$. The local excitation intensity enhancement $\eta_{exc}$ is deduced from numerical simulations by averaging the electric field intensity in the lower half part of the aperture \cite{Blair10,Blair07}. The gain in collection efficiency is assumed to be equal to $\eta_{\kappa}=1.5$ for both aluminum and gold 190~nm ZMWs, following earlier works on aperture directivity \cite{Davy08,Aouani11}. The gain in quantum efficiency $\eta_{\phi} = \eta_{rad} / \eta_{tot}$ can be rewritten as the ratio of the gain in radiative rate $\eta_{rad}$ over the total (radiative + nonradiative) decay rate enhancement $\eta_{tot}$. The total decay rate enhancement $\eta_{tot}$ corresponds to the lifetime reduction as measured in Fig.~\ref{FigEnhFRET}a. The radiative rate enhancement can then be computed from all previous measurements as $\eta_{rad}= \eta_F \eta_{tot}/ (\eta_{exc} \eta_{\kappa})$. The non-radiative rate (without FRET) enhancement is computed following the relation $\eta_{tot}= \phi \eta_{rad}+ (1-\phi) \eta_{nrad}$ and the known quantum yield $\phi=0.8$ for Atto550 in confocal case.

\section*{Acknowledgment}
The research leading to these results has received funding from the European Commission's Seventh Framework Programme (FP7-ICT-2011-7) under grant agreement ERC StG 278242 (ExtendFRET). PG is on leave from Institute for Space Sciences, Bucharest-M\u agurele RO-077125, Romania.

\newpage

\section*{Supporting Information}

\subsection*{1. The two FRET measurement methods converge towards similar results}

\begin{figure}[h!]
\begin{center}
\includegraphics[scale=1.2]{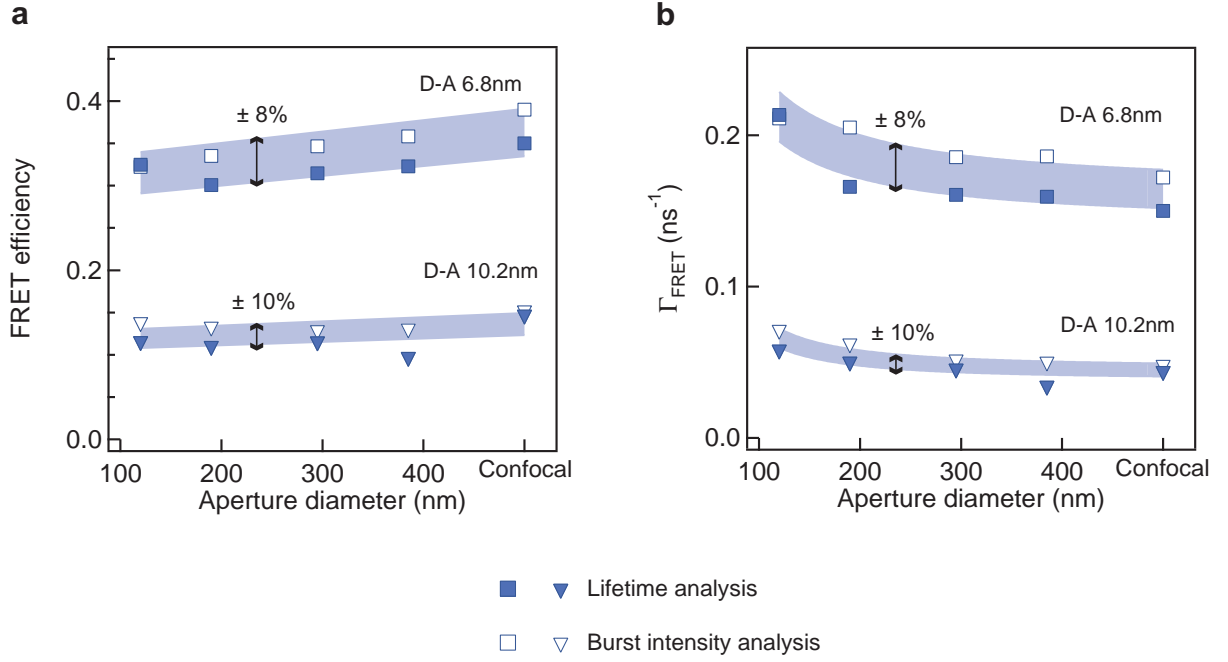}
\caption{Comparison of the results obtained from analysis of the donor fluorescence lifetime (filled markers) and the acceptor/donor burst intensities (empty markers). Both methods converge remarkably towards similar results for both the FRET efficiency (a) and the FRET rate (b). The blue-grey solid regions indicate deviation of 8\% from the average for D-A distances of 6.8~nm, and 10\% for D-A distances of 10.2~nm. To extract the FRET rate from the burst intensity analysis, we use $\Gamma_{FRET}= \Gamma_{D} E_{FRET}/(1-E_{FRET})$ where $\Gamma_{D}=1/\tau_{D}$ is the donor decay rate obtained from time-correlated lifetime measurements on the isolated donor case. The results shown here are for aluminum ZMWs. Similar results are obtained for gold nanoapertures \cite{Ghenuche14}.}
\end{center}
\end{figure}

\newpage

\subsection*{2. FCS quantifies the diffusion properties in ZMWs and the fluorescence brightness enhancement}

Fluorescence correlation spectroscopy (FCS) is used to quantify the number of molecules $N$ contributing to the detected fluorescence signal and the mean translational diffusion time $\tau_d$ for the fluorophores to cross the detection volume \cite{Maiti,Fluobouquin}. FCS performs a statistical analysis of the temporal fluctuations affecting the fluorescence intensity  by computing the second order correlation of the fluorescence intensity time trace $G^{(2)}(\tau) = \langle \delta F(t). \delta F(t+\tau) \rangle / \langle F(t) \rangle ^2$, where $\delta F(t) = F(t) - \langle F \rangle $ is the fluctuation of the time-dependent fluorescence signal $F(t)$, $\tau$ the delay (lag) time, and $\langle . \rangle$ stands for time averaging. The analysis of the FCS correlation function is based on a three dimensional Brownian diffusion model \cite{Maiti,Fluobouquin}:
\begin{equation}\label{Eq:diffFCS}
   G^{(2)}(\tau) = \frac{1}{N}\, \left( 1 - \frac{\langle B \rangle}{\langle F \rangle}\right)^2 \, \left[1 + \frac{T}{1-T} \exp \left(-\frac{\tau}{\tau_{T}} \right) \right]   \frac{1}{(1+\tau/\tau_d)\sqrt{1+(1/\kappa)^2 \, \tau/\tau_d}}
\end{equation}
\noindent where $N$ is the total number of molecules, $\langle F \rangle$ the average intensity, $\langle B \rangle$ the background noise, $T$ the fraction of dyes in the dark state, $\tau_{T}$ the dark state blinking time, $\tau_d$ the mean diffusion time and $\kappa$ the aspect ratio of the axial to transversal dimensions of the analysis volume. The background noise $\langle B \rangle$ originates mainly from the back-reflected laser light and from metal photoluminescence. At 40~$\mu$W excitation power, it is typically of 0.5~kHz, which is quite negligible as compared to the typical fluorescence brightness per molecule. Triplet blinking is quite weak for the Atto550 dye. The triplet fraction $T$ converges to a value of 9\% for the confocal reference, and vanishes for the FCS traces in ZMWs. Therefore, we decided to set $T=0$ for the FCS analysis in ZMWs. The shape parameter $\kappa$ is fixed to $\kappa=5$ for the confocal microscope from the PSF calibration, and kept as a free parameter for the fits in the ZMWs. As already noted in our previous works \cite{Wenger08,Davy08}, this parameter converges towards values close to 1 in ZMWs.

To compute the FCS correlation, the fluorescence intensity trace is recorded on the Hydraharp400 time-correlated single photon counting module (Picoquant GmbH) in time-tagged time-resolved (TTTR) mode. Each trace duration lasts typically 200s. The correlation is then computed using the software Symphotime 64 (Picoquant GmbH) with lag times ranging from 1~$\mu$s to 1~s. To avoid artifacts on the autocorrelation curve at sub-microsecond lag times related to photodiode afterpulsing, we implement the FLCS background correction as described in \cite{EnderleinFLCS}. The FCS curves are fitted using the model of Eq.~(1) and Levenberg-Marquard optimization. This procedure quantifies the number of molecules $N$ and their mean translational diffusion time $\tau_d$. The brightness per molecule is then computed as $Q = (\langle F \rangle-\langle B \rangle)/N $. Comparing the brightness in the ZMW and the confocal reference then quantifies the fluorescence brighness enhancement factor $\eta_F = Q_{ZMW} / Q_{Confoc}$. Figure S2 presents typical FCS correlation data recorded in confocal diffraction-limited mode and in 120~nm and 190~nm aluminum nanoapertures. The fit results are summarized in Tab.~S1.

\newpage

\begin{figure}[h!]
\begin{center}
\includegraphics{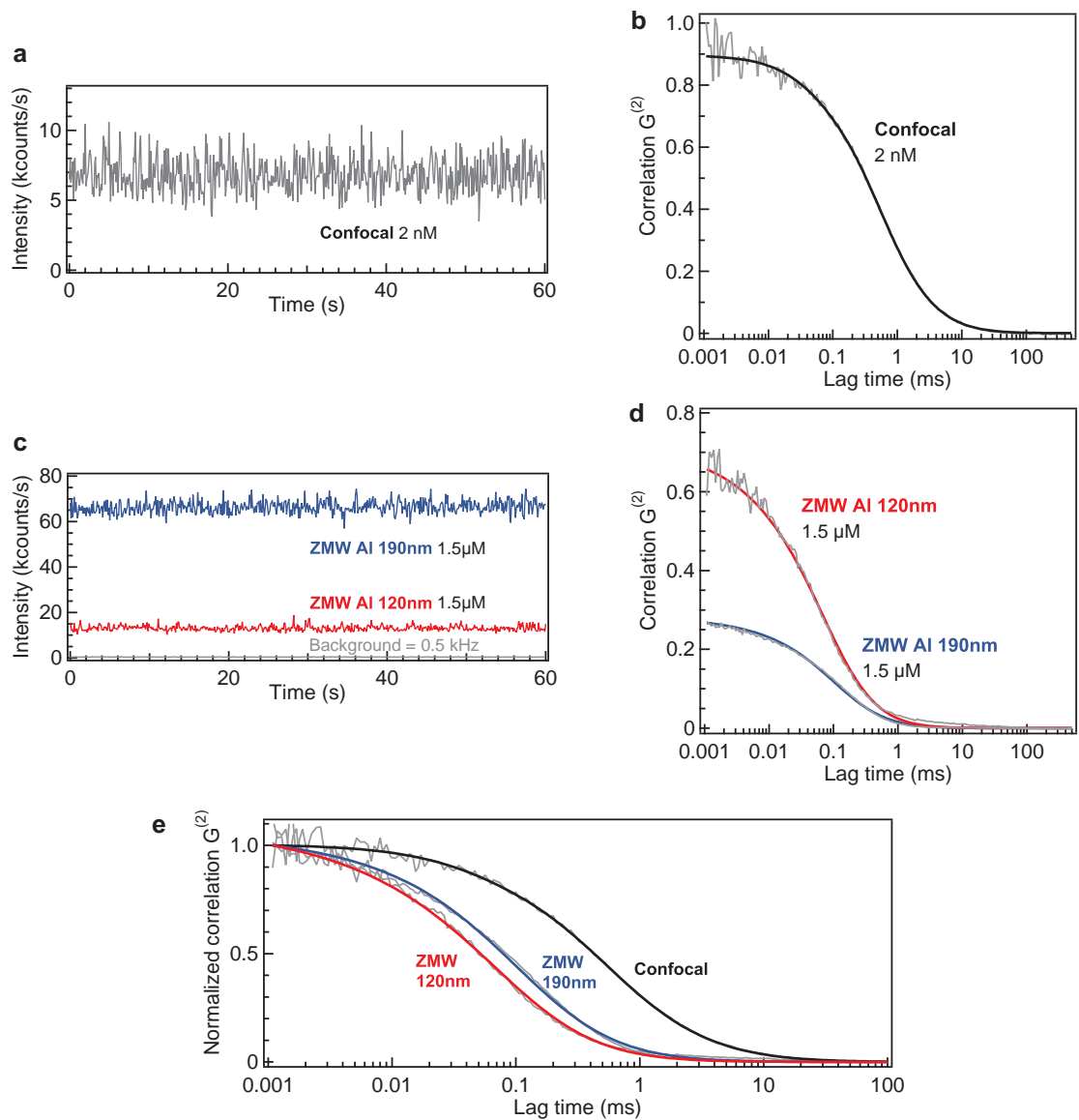}
\caption{Fluorescence intensity time trace and corresponding FCS correlation function for the confocal reference (a,b) and nanoapertures milled in aluminum (c,d). Thick lines in (b,d) are numerical fits using Eq.~(1) model. The fit results are summarized in Tab.~S1. The results for 190 and 120~nm ZMWs use the same concentration to enable a direct comparison between the fluorescence signals and detection volumes. (e) Normalized correlation functions showing the reduction of diffusion time across the detection volume as the ZMW diameter is reduced.}
\end{center}
\end{figure}

\newpage

\begin{table}[h!]
\centering
\begin{tabular}{|cc|c|c|c|}\hline
 & & Confocal & ZMW 190nm & ZMW 120nm\\ \hline
Number of molecules & N  & 1.2 & 3.5 & 1.3 \\
Diffusion time & $\tau_d$ ($\mu$s) & 540 & 170 & 140 \\
Triplet fraction & T  & 0.09 & 0 & 0 \\
Blinking time & $\tau_T$ ($\mu$s)  & 40 & - & - \\
Shape ratio & $\kappa$  & 5 & 1.5 & 1 \\\hline
Average intensity & $\langle F \rangle$ (kHz) & 6.8 & 66.6 & 13.2 \\
Background noise & $\langle B \rangle$ (kHz) & 0.1 & 0.5 & 0.5 \\
Brightness per molecule & Q (kHz)  & 5.6 & 18.9 & 9.8 \\
Fluorescence enhancement & $\eta_F$ & - & 3.3 & 1.7 \\\hline
Concentration used & C (nM)  & 2 & 1500 & 1500 \\
Detection volume & $V_{eff}$ (fL)  & 1 & 0.0038 & 0.0014 \\
Volume reduction & $R_{vol}$  & - & 265 & 710 \\\hline
\end{tabular}
\caption{The parameters obtained by fitting the FCS correlograms displayed in Fig. S2b,d are used to quantify the fluorescence brightness enhancement and the detection volume reduction.}\label{Tab1}
\end{table}

\newpage

\subsection*{3. The FRET results are invariant upon 4x concentration change}

\begin{figure}[h!]
\begin{center}
\includegraphics{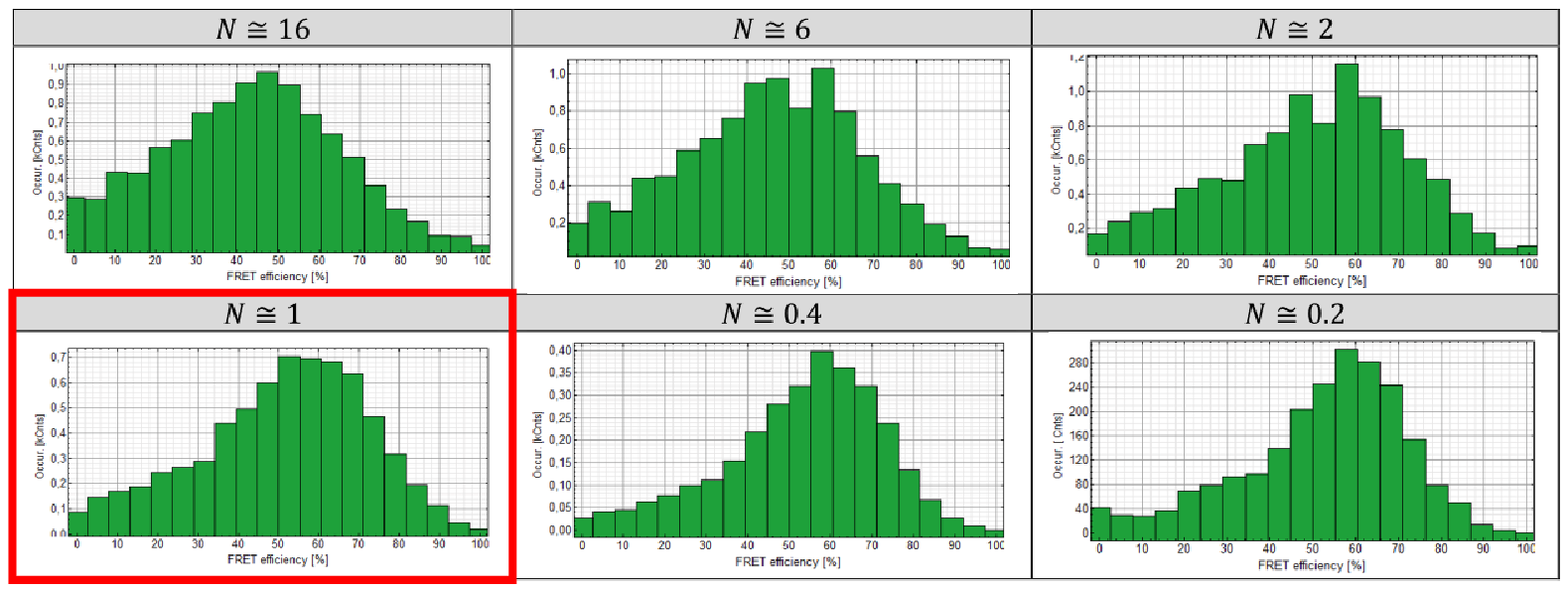}
\caption{FRET histograms as function of the average number $N$ of FRET pairs in the detection volume, as determined by FCS. The case $N\simeq1$ is the one used for our experiments. No significant variation on the FRET histogram is seen when the concentration is raised up to 4-6 times. At very high concentrations ($N\simeq16$) deviations occur due to spatial averaging, yet this condition is always avoided in our study. The threshold for burst detection was set to the minimum value (mean of the trace) to better reveal small variations in the histograms. Consequently, the histogram widths are slightly larger here.}
\end{center}
\end{figure}

\newpage

\subsection*{4. Comparison between FRET histograms}

\begin{figure}[h!]
\begin{center}
\includegraphics[scale=0.94]{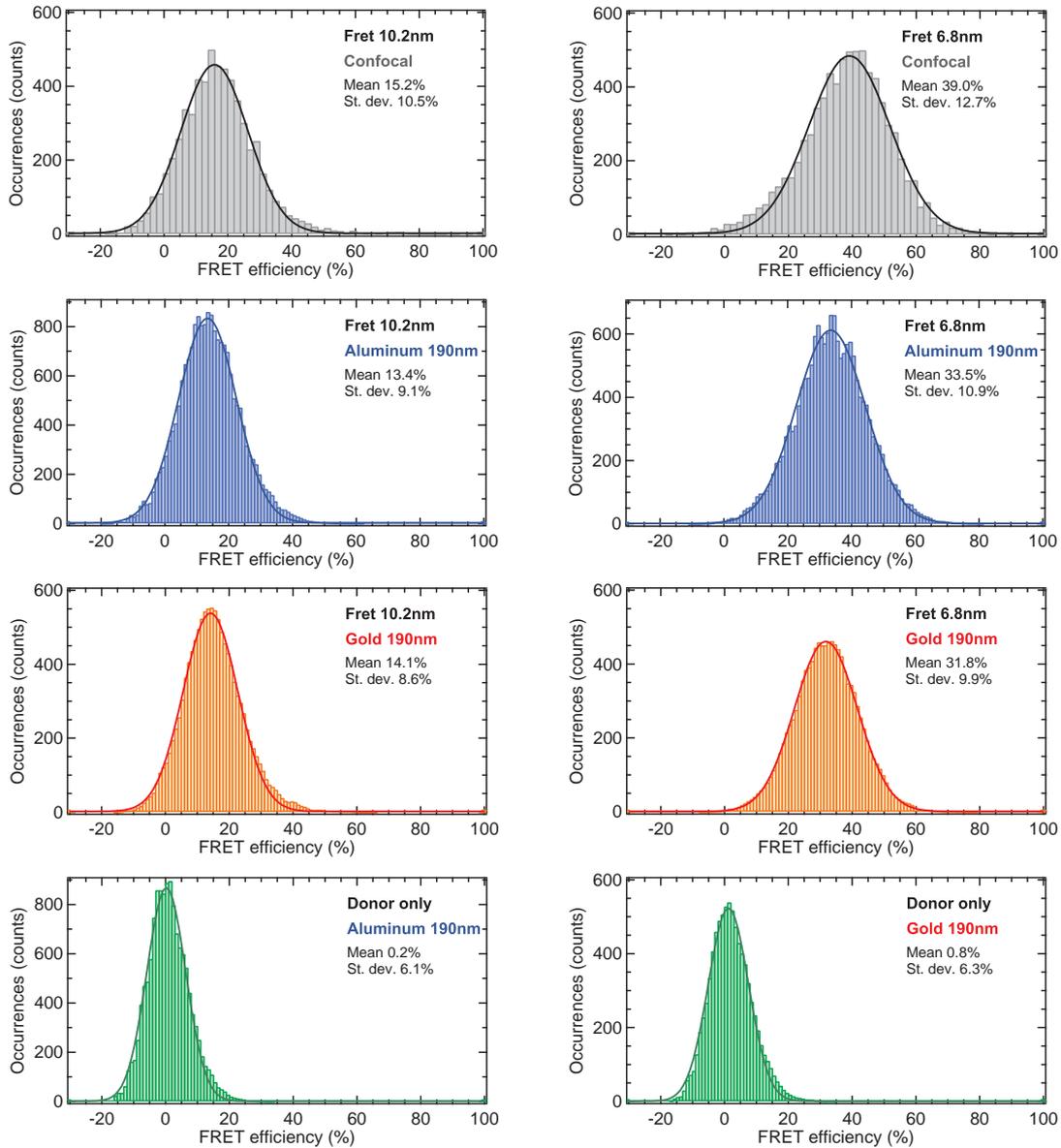}
\caption{Non-normalized experimental FRET histograms together with Gaussian fits used to determine the center FRET efficiency and the standard deviation. Events with apparent transfer efficiency below zero are also shown. We also display the histograms obtained on the DNA samples containing only the donor fluorophore to provide a reference for the zero FRET case.}
\end{center}
\end{figure}

\newpage

\subsection*{5. FRET histogram widths}

\begin{figure}[h!]
\begin{center}
\includegraphics[scale=0.7]{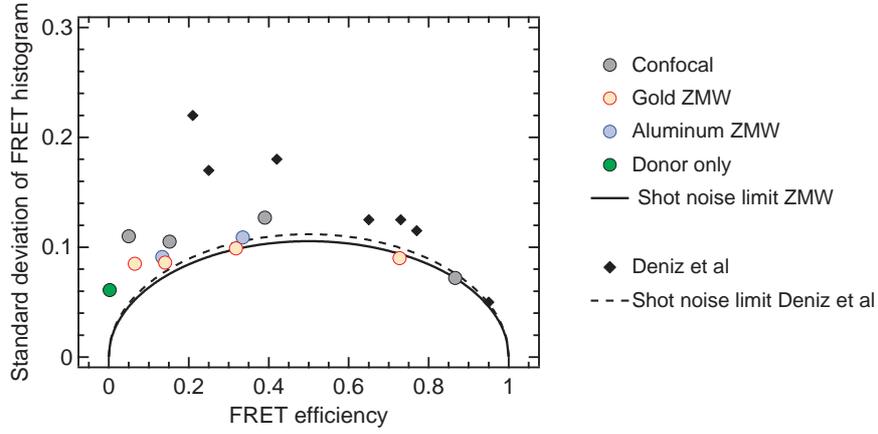}
\caption{Standard deviation extracted from the Gaussian fits of the FRET histograms as function of the mean FRET efficiencies. For comparison to our data, we also report the values taken from Deniz et al Ref.~\cite{Deniz99}. Additional data on gold ZMW and confocal measurements are also taken from our previous work \cite{Ghenuche14}. The shot noise limit is computed following the common simple approach $\sigma_E = \sqrt{E(1-E)/T}$, as found in \cite{Deniz99,Deniz01}. $E$ is the average FRET efficiency, $T$ is the threshold level which is $T=20$ for Ref.~\cite{Deniz99} and $T=22.2$ for our data.}
\end{center}
\end{figure}

\newpage

\subsection*{6. Fluorescence decay traces in gold have a supplementary fast decay related to gold photoluminescence}

\begin{figure}[h!]
\begin{center}
\includegraphics{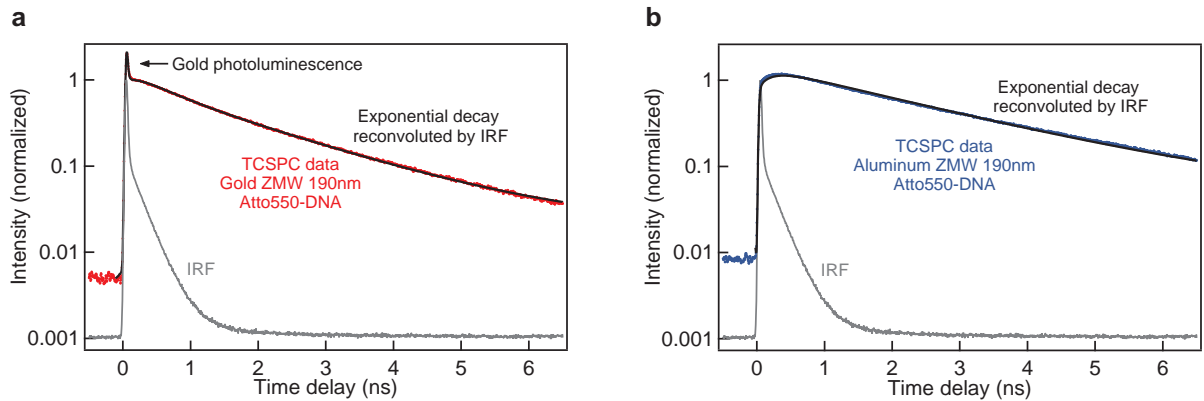}
\caption{Normalized donor fluorescence decay traces for a 190~nm aperture milled in gold (a) or aluminum (b). Black lines represent the results of the exponential fit model reconvoluted by the instrument response function (IRF, grey trace). In the case of gold, a fast ($\sim 5$~ps, limited by the resolution of our instrument) supplementary decay is visible, that we relate to the photoluminescence of gold excited at 550~nm.}
\end{center}
\end{figure}

\newpage

\subsection*{7. Table of donor fluorescence lifetimes measured in ZMWs}

\begin{table}[h!]
\centering
\begin{tabular}{|c|c|c|c|c|}\hline
  & Diameter (nm) & Isolated donor & FRET 6.8~nm & FRET 10.2~nm  \\
   &  & $\tau_D$ (ns) & $\tau_{DA}$ (ns) & $\tau_{DA}$ (ns) \\ \hline
  Confocal & - & 3.67 & 2.44 & 3.13 \\ \hline
  Aluminum  & 120  & 2.25  & 1.52  & 1.99  \\
    & 190  &  2.46 & 1.72  & 2.19  \\
    & 295  &  2.86 & 1.96  & 2.53  \\
    & 385  &  3.01 & 2.05  & 2.72  \\ \hline
 Gold & 150  & 1.30  & 0.90  & 1.12  \\
    & 190  & 1.65  & 1.09  & 1.42  \\
    &  230 & 1.97  & 1.31  & 1.67  \\
    & 280  & 2.08  & 1.38  & 1.83  \\
    &  330 & 2.26  & 1.46  & 1.92  \\
    & 380  & 3.02  & 1.86  & 2.50  \\ \hline
\end{tabular}
\caption{Donor fluorescence lifetimes displayed in Fig. 2b,c.}\label{Tab1}
\end{table}

\newpage

\subsection*{8. Side by side comparison of enhancement factors for 190nm nanoapertures}

\begin{figure}[h!]
\begin{center}
\includegraphics[width=14cm]{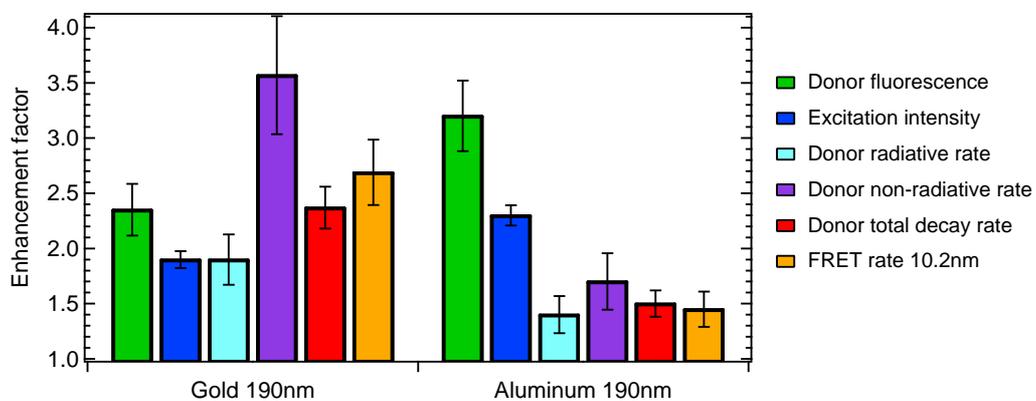}
\caption{Summary of the enhancement factors for 190~nm ZMWs milled in aluminum or gold. The gain in collection efficiency is similar for both metals and amounts to 1.5x.} \label{FigComp}
\end{center}
\end{figure}

\newpage

\subsection*{9. Choosing the most appropriate metal: a figure of merit as function of spectral range}

\begin{figure}[h!]
\begin{center}
\includegraphics{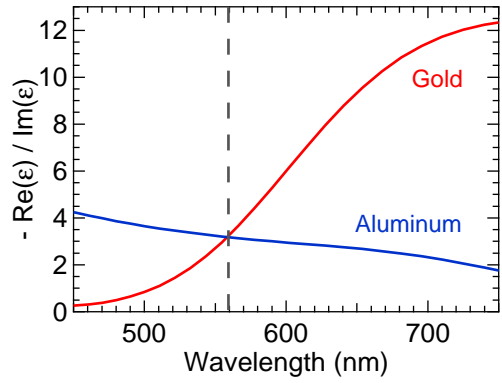}
\caption{Figure of merit for plasmonic metals defined as the ratio of the absolute value of the real part over the imaginary part of the complex permittivity of gold (red) and aluminum (blue) \cite{West,Jackson}. The metal permittivity data is extracted from \cite{Rakic,RefractiveIndex}. The dashed line indicates the crossing between gold and aluminum near 560~nm.} \label{FigLoss}
\end{center}
\end{figure}

\newpage


\end{document}